\documentclass[draft]{agujournal2019}
\usepackage{url} %this package should fix any errors with URLs in refs.
\usepackage{lineno}
\usepackage[inline]{trackchanges} %for better track changes. finalnew option will compile document with changes incorporated.
\usepackage{soul}
\draftfalse

\journalname{JGR: Planets}

\usepackage{pdflscape,epstopdf,textgreek,tabularx,colortbl}
\usepackage[shortlabels]{enumitem}
\newcolumntype{L}{>{\raggedright\arraybackslash}X}
\begin{document}

%% ------------------------------------------------------------------------ %%
%
%  TITLE
%
%% ------------------------------------------------------------------------ %%

\title{Meridional variations of C\textsubscript{2}H\textsubscript{2} in Jupiter's stratosphere from Juno UVS observations}

%% ------------------------------------------------------------------------ %%
%
%  AUTHORS AND AFFILIATIONS
%
%% ------------------------------------------------------------------------ %%

\authors{Rohini S. Giles\affil{1}, Thomas K. Greathouse\affil{1}, Vincent Hue\affil{1}, G. Randall Gladstone\affil{1,2}, Henrik Melin\affil{3}, Leigh N. Fletcher\affil{3}, Patrick G. J. Irwin\affil{4}, Joshua A. Kammer\affil{1}, Maarten H. Versteeg\affil{1}, Bertrand Bonfond\affil{5}, Denis C. Grodent\affil{5}, Scott J. Bolton\affil{1}, and Steven M. Levin\affil{6}}

\affiliation{1}{Space Science and Engineering Division, Southwest Research Institute, San Antonio, Texas, USA}
\affiliation{2}{Department of Physics and Astronomy, University of Texas at San Antonio, San Antonio, Texas, USA}
\affiliation{3}{School of Physics \& Astronomy, University of Leicester, Leicester, United Kingdom}
\affiliation{4}{Department of Physics, University of Oxford, Oxford, United Kingdom}
\affiliation{5}{Laboratoire de Physique Atmosphérique et Planétaire, STAR Institute, Université de Liège, Liège, Belgium}
\affiliation{6}{Jet Propulsion Laboratory, Pasadena, California, USA}

%% ------------------------------------------------------------------------ %%
%
%  KEY POINTS
%
%% ------------------------------------------------------------------------ %%

%\begin{keypoints}
%\item .
%\end{keypoints}

%% ------------------------------------------------------------------------ %%
%
%  ABSTRACT
%
%% ------------------------------------------------------------------------ %%

\begin{abstract}

The UVS instrument on the Juno mission records far-ultraviolet reflected sunlight from Jupiter. These spectra are sensitive to the abundances of chemical species in the upper atmosphere and to the distribution of the stratospheric haze layer. We combine observations from the first 30 perijoves of the mission in order to study the meridional distribution of acetylene (C\textsubscript{2}H\textsubscript{2}) in Jupiter's stratosphere. %These are the first full-planet meridional retrievals of C\textsubscript{2}H\textsubscript{2} from ultraviolet spectra.
We find that the abundance of C\textsubscript{2}H\textsubscript{2} decreases towards the poles by a factor of 2--4, in agreement with previous analyses of mid-infrared spectra. This result is expected from insolation rates: near the equator, the UV solar flux is higher, allowing more C\textsubscript{2}H\textsubscript{2} to be generated from the UV photolysis of CH\textsubscript{4}. The decrease in abundance towards the poles suggests that horizontal mixing rates are not rapid enough to homogenize the latitudinal distribution. 
    
\end{abstract}

\section*{Plain Language Summary}

The UVS instrument on the Juno mission to Jupiter is primarily used to study the planet's ultraviolet auroras, but also records reflected sunlight from the planet's upper atmosphere. These ultraviolet reflected sunlight observations can be used to measure the abundances of different gases in Jupiter's stratosphere. In this paper, we focus on one prominent molecule, acetylene, and study how its abundance varies with latitude. We find that its abundance decreases towards Jupiter's poles, which agrees with previous results obtained from studying the same molecule with infrared observations.

%% ------------------------------------------------------------------------ %%
%
%  INTRODUCTION
%
%% ------------------------------------------------------------------------ %%

\section{Introduction}

Hydrocarbons play a significant role in Jupiter's stratosphere. Methane (CH\textsubscript{4}) is the most abundant spectroscopically absorbing gas in the stratosphere and its photolysis by UV photons leads to a rich hydrocarbon photochemistry \cite{moses04}. The most abundant products of CH\textsubscript{4} photolysis are ethane (C\textsubscript{2}H\textsubscript{6}) and acetylene (C\textsubscript{2}H\textsubscript{2}), the latter of which is the focus of this paper. C\textsubscript{2}H\textsubscript{6} and C\textsubscript{2}H\textsubscript{2} were first detected in Jupiter's stratosphere by \citeA{ridgway74} using groundbased observations at 10 \textmu m, and this was followed by the subsequent detections of other hydrocarbons including C\textsubscript{2}H\textsubscript{4}, C\textsubscript{3}H\textsubscript{4}, and C\textsubscript{6}H\textsubscript{6}~\cite{kim85}.

As disequilibrium species, these hydrocarbons act as tracers of the atmospheric circulation and chemistry in Jupiter's stratosphere. Spatial variability in the abundance of a given disequilibrium species can provide valuable constraints in understanding how production rates and mixing rates vary across the planet. Non-auroral latitudinal variations can provide information about horizontal eddy diffusion rates~\cite{lellouch06,nixon07,hue18b}, while spatial variability within the auroral region can tell us about auroral dynamics and enhanced ion-related chemistry due to auroral precipitation ~\cite{wong00,sinclair19,sinclair20}.

The focus of this paper is latitudinal variations in C\textsubscript{2}H\textsubscript{2} outside of the auroral regions. Since its first detection by \citeA{ridgway74}, numerous studies have constrained Jupiter's stratospheric C\textsubscript{2}H\textsubscript{2} abundance using observations from both the infrared \cite<e.g.,>{drossart86, sada98,fouchet00c} and the ultraviolet \cite<e.g.,>{gladstone83}. The first spatially-resolved full-planet retrievals of C\textsubscript{2}H\textsubscript{2} abundance were carried out by \citeA{kunde04} and \citeA{nixon07}, using infrared observations from the Cassini CIRS instrument and similar analyses were later carried out using Voyager data \cite{nixon10,zhang13b}. \citeA{nixon07} found that the 5-mbar C\textsubscript{2}H\textsubscript{2} abundance peaked at 20$^{\circ}$N and then decreased towards the poles by a factor of $\sim$4. Similar latitudinal trends have also been observed using groundbased mid-infrared spectroscopic observations \cite{fletcher16,melin18}.

As discussed in \citeA{nixon07}, a decrease in C\textsubscript{2}H\textsubscript{2} abundance towards the poles can be understood by considering the lifetime of C\textsubscript{2}H\textsubscript{2} in Jupiter's stratosphere. At 5 mbar, C\textsubscript{2}H\textsubscript{2} has a chemical lifetime of $3\times10^{7}$ s, which is shorter than the expected horizontal mixing timescales of \textgreater1 Jovian year ($4\times10^{8}$ s). The incident UV flux is greater at the equator than at high latitude, causing more  C\textsubscript{2}H\textsubscript{2} to be produced at low latitude and the short lifetime means that it is not transported to the poles rapidly enough to homogenize the latitudinal distribution. Assuming the C\textsubscript{2}H\textsubscript{2} chemistry is well known, its meridional distribution can therefore be used to constrain the horizontal mixing timescale and a comparison of the meridional distribution at different pressure levels can constrain how the horizontal mixing changes with altitude. 

In this paper, we present the first full-planet meridional retrievals of C\textsubscript{2}H\textsubscript{2} from ultraviolet reflected sunlight observations of Jupiter, using data from the UVS instrument on the Juno spacecraft. This builds on previous work from \citeA{melin20} who retrieved low-latitude abundances from Cassini UVIS data, which has a lower spatial resolution but higher spectral resolution than the Juno UVS observations. In Section~\ref{sec:observations} we describe the observations made with the UVS instrument on the Juno spacecraft. In Section~\ref{sec:analysis} we describe the meridional retrieval of the stratospheric C\textsubscript{2}H\textsubscript{2} abundance and in Section~\ref{sec:discussion} we compare this distribution with previous meridional distributions from infrared observations.

%% ------------------------------------------------------------------------ %%
%
%  OBSERVATIONS
%
%% ------------------------------------------------------------------------ %%

\section{Observations}
\label{sec:observations}

\subsection{Juno UVS}

The Ultraviolet Spectrograph \cite<UVS,>{gladstone17} is a far-ultraviolet imaging spectrograph on NASA's Juno mission, which arrived at Jupiter in July 2016 and is currently in a highly elliptical 53-day polar orbit around the planet \cite{bolton17}. Observations of Jupiter are obtained for several hours on either side of each perijove (PJ), the closest point of approach in the orbit. The primary scientific goal of UVS is to study Jupiter's auroral emissions and the instrument's spectral range of 68--210 nm was designed to cover several important H and H\textsubscript{2} emission bands. However, the longer wavelengths covered by UVS extend into a spectral region that is dominated by reflected sunlight; this part of the spectrum is sensitive to hydrocarbons and aerosols \cite{gladstone17} and will be the focus of this paper. At \textgreater200 nm, the UVS sensitivity drops off significantly, increasing the uncertainty on the radiometric calibration. In this paper, we focus on the 164--200 nm segment of the spectrum. At these wavelengths, ultraviolet reflected photons are primarily reflected from the 5--50 pressure level~\cite{melin20} and the spectrum is therefore sensitive to molecular absorbers in this region of the atmosphere or higher. 

UVS has a dog-bone shaped slit with a total length of 7.2$^{\circ}$. The outer wide segments of the slit have a spectral resolution of 2--3 nm and the central narrow segment has a spectral resolution of $\sim$1.3 nm. In this paper, we only use data from the wide segments of the slit, as the signal to noise ratio is significantly better. As the Juno spacecraft spins with a period of $\sim$30 s, the field-of-view defined by the UVS slit sweeps across a swath of Jupiter. Ultraviolet photon detections are recorded in a pixel-list format, containing information about the wavelength of the photon, the position along the slit and the precise time of the detection. The position along the slit can be combined with information about the orientation of the spacecraft at the time of observation in order to map the photon to a latitude and longitude on the planet. This allows spectral image cubes of the planet to be built up. 

\subsection{Reflected sunlight spectra}

As Juno is in a polar orbit around Jupiter, the majority of the UVS observations cover Jupiter's polar regions. An analysis of the UVS reflected sunlight observations of Jupiter's poles will be the focus of a future study; in this paper, we focus on meridional trends and limit our analysis to regions of the planet without auroral emissions, which extend into the reflected sunlight part of the spectrum and complicate radiative transfer modeling.

In order to obtain sufficient signal-to-noise at all latitudes, we combine wide-slit dayside observations from the first 30 perijoves of the mission (August 2016 - November 2020) and bin them into 10$^{\circ}$-wide latitude bins. We use all data from the main perijove observing sequence, which is typically $\pm$5 hours from the time of closest approach. In order to avoid the auroral regions of the planet, we exclude longitudes and latitudes that fall poleward of a boundary oval, set at 5000 km equatorward of the Io footprint reference oval \cite{bonfond17b}. Figure S1 in the Supporting Information shows maps of these excluded auroral regions.

For the combined observations in each latitude bin, we produce a 2-dimensional histogram of emission angle and solar zenith angle in order to determine the geometric coverage. Figure S2 in the Supporting Information shows these histograms for two example latitude bins. In order to easily compare different latitudes, we looked for a combination of angles that were available at all latitudes. We found that all latitude bins included observations obtained with an emission angle of 40--50$^{\circ}$ and a solar zenith angle of 70--80$^{\circ}$; this combination is shown by the red boxes in Figure S2. For each latitude bin, data fulfilling these angle criteria were averaged together, producing a set of representative spectra at each latitude, each with a comparable observing geometry. These representative spectra combine observations from many perijoves. 

\begin{figure}
\centering
\includegraphics[width=10cm]{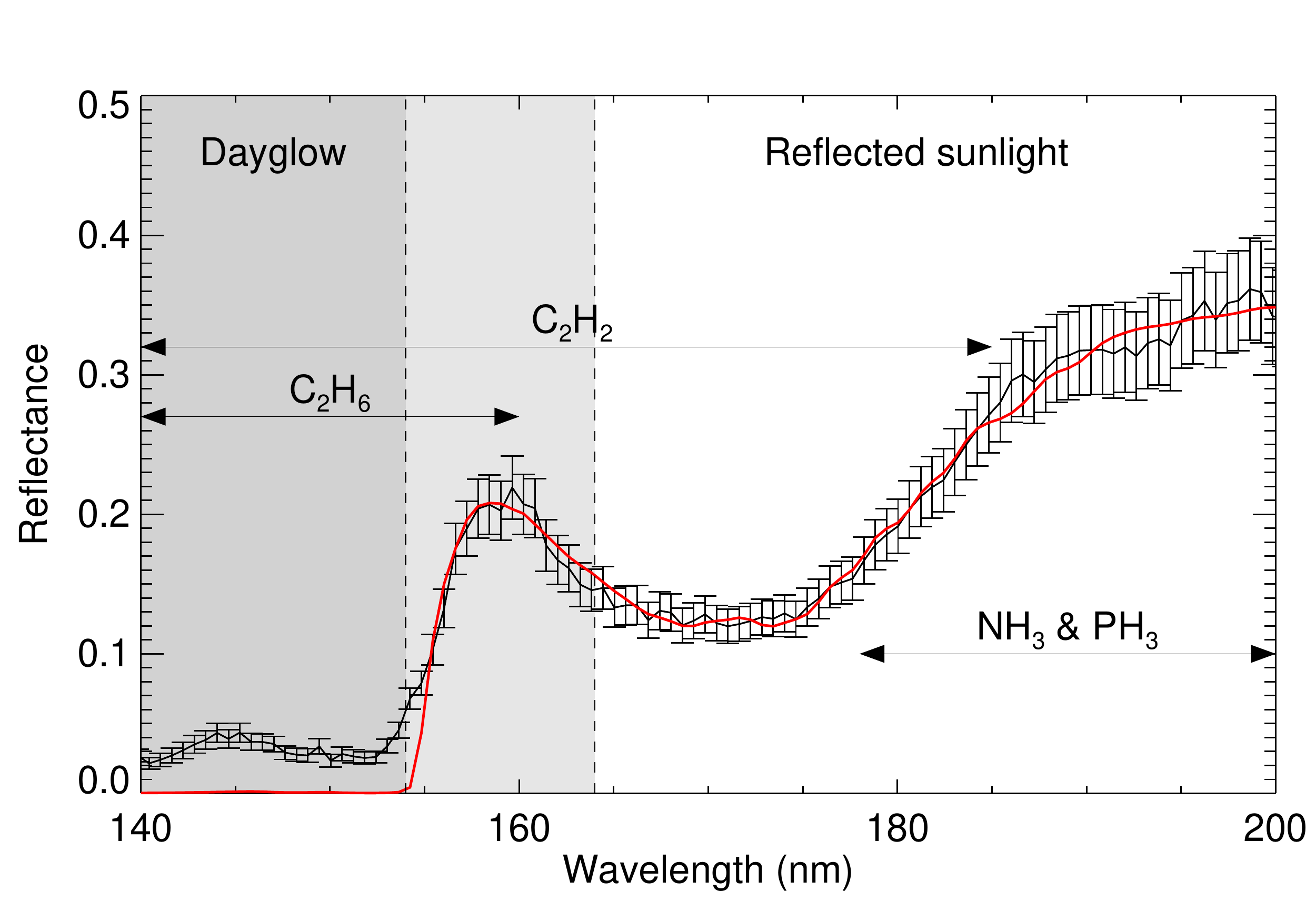}
\caption{Jupiter's equatorial reflectance spectrum (corrected for the solar zenith angle) observed by Juno UVS (black) alongside the best-fit model NEMESIS spectrum (red). The primary molecular absorber in this spectral range is C\textsubscript{2}H\textsubscript{2} but there are also smaller contributions from C\textsubscript{2}H\textsubscript{6}, NH\textsubscript{3} and PH\textsubscript{3}. At \textless 154 nm, the spectrum is dominated by dayglow and at \textgreater 164 nm, the spectrum is dominated by reflected sunlight. In this paper, we focus on the 164--200 nm segment of the UVS spectrum.}
\label{fig:spectrum}
\end{figure}

An example Juno UVS spectrum from the latitude bin centered on the equator is shown in black in Figure~\ref{fig:spectrum}. This spectrum is shown in terms of the planet's reflectance i.e., the observed spectrum has been divided by the incident solar spectrum, corrected for the solar zenith angle. UVS observations have background radiation noise due to the penetrating high-energy electrons in the environment around Jupiter~\cite{kammer19}. We estimate the radiation noise level from the average count rate at \textless80 nm (where we expect the counts to all be due to radiation) and this has been subtracted from the spectrum. The error bars are calculated by adding in quadrature the radiation noise and the shot noise (a signal-to-noise ratio of $\sqrt{N}$, where $N$ is the number of photon counts recorded before radiation noise subtraction). The red line in Figure~\ref{fig:spectrum} shows the best-fit model reflected sunlight spectrum obtained with the NEMESIS radiative transfer and retrieval tool and is discussed in Section~\ref{sec:analysis}.

At \textless154 nm, the Jovian spectrum is dominated by H\textsubscript{2} Lyman band emission from dayglow, rather than reflected sunlight~\cite{morrissey95}. At 154--164 nm, both dayglow and reflected sunlight contribute to the shape of the spectrum and by \textgreater164 nm, reflected sunlight is the dominant component. The dominant stratospheric species that affects the spectral shape in the 140--200 nm wavelength range is C\textsubscript{2}H\textsubscript{2}, which is the focus of this study. C\textsubscript{2}H\textsubscript{2} has a peak in absorption at $\sim$172 nm, causing the broad dip seen in Figure~\ref{fig:spectrum}. The sharp drop at the short-wavelength edge of the spectrum is primarily due to C\textsubscript{2}H\textsubscript{6} absorption, although C\textsubscript{2}H\textsubscript{2} also plays a role. The peak at 160 nm is primarily caused by the absorption cross-section of these two gases, although there is also a peak in the H\textsubscript{2} dayglow emission at this wavelength which can further increase the radiance. Due to this, we restrict our radiative transfer modeling in the following sections to the 164--200 nm segment of the UVS spectrum. 

C\textsubscript{2}H\textsubscript{4} and C\textsubscript{4}H\textsubscript{2} also absorb light in this part of the ultraviolet spectrum, but at the abundances in which they are present in the stratosphere, their impact on the spectral shape is minor. The other two absorbing species of note are NH\textsubscript{3} and PH\textsubscript{3}; these gases are not normally present in Jupiter's stratosphere, although stratospheric NH\textsubscript{3} was observed in the aftermath of the Shoemaker–Levy 9 impact on Jupiter \cite{kostiuk96} and the 2009 impactor \cite{fletcher11d}. However, their presence in the upper troposphere can affect the spectral shape, particularly at the longer wavelengths where there is less absorption from stratospheric species. Forward models showing the individual impacts of these molecular species are presented in \citeA{melin20}.

\begin{figure}
\centering
\includegraphics[width=9.5cm]{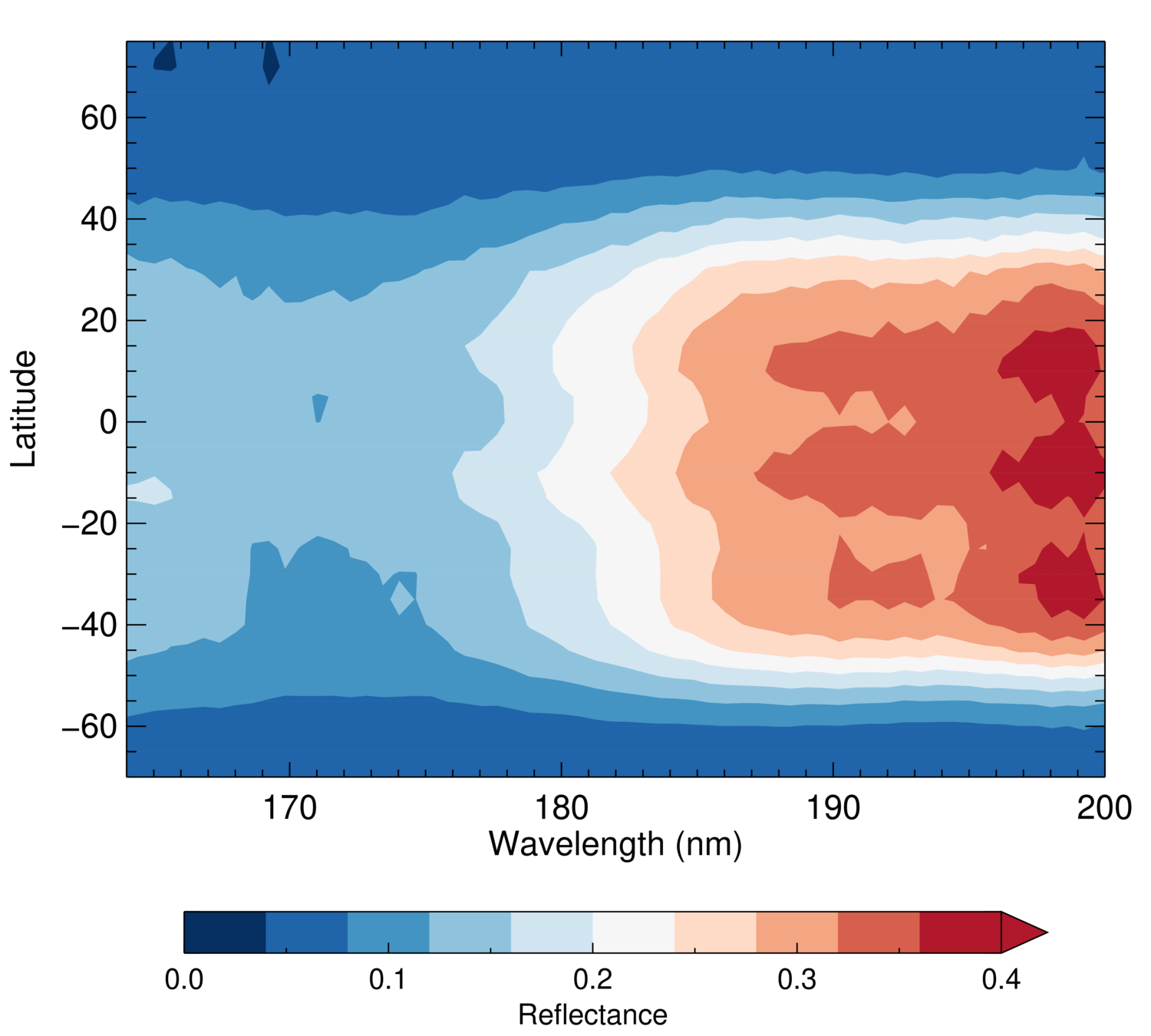}
\caption{Contour plot of the UVS reflected sunlight spectra (corrected for the solar zenith angles) as a function of latitude. At high latitudes, the average reflectance is lower and the spectra are flatter.}
\label{fig:contour}
\end{figure}

Figure~\ref{fig:spectrum} showed an example spectrum from the latitude bin centered on Jupiter's equator. Inspection of the spectra from all latitude bins shows that the spectra vary considerably. Firstly, the average reflectance decreases towards the poles, which is due to the increasing opacity of the stratospheric haze. In addition to the average reflectance changing, we also find that the spectral shape at 164--200 nm changes. These effects are shown in Figure~\ref{fig:contour}, which presents a contour plot of the spectra as a function of latitude. At higher latitudes, Figure~\ref{fig:contour} shows a lower average reflectance. These high latitude spectra are also relatively flat across the entire 164--200 nm spectral range, while spectra from near the equator show a broad minimum at $\sim$172 nm. As C\textsubscript{2}H\textsubscript{2} has a peak in its absorption cross section at $\sim$172 nm, this suggests that the variability in spectral shape could be due to decreasing C\textsubscript{2}H\textsubscript{2} abundance at high latitudes. This is explored using a radiative transfer and retrieval code in Section~\ref{sec:analysis}.

%% ------------------------------------------------------------------------ %%
%
%  ANALYSIS
%
%% ------------------------------------------------------------------------ %%

\section{Analysis}
\label{sec:analysis}

\subsection{Radiative transfer and retrieval code}
\label{sec:radiative_transfer}

In order to determine the C\textsubscript{2}H\textsubscript{2} abundance as a function of latitude, we analyzed the Juno UVS spectra using the NEMESIS radiative transfer and retrieval tool \cite{irwin08}. NEMESIS was originally developed in order to analyze infrared observations from Cassini CIRS, but was later expanded to cover visible and sub-mm wavelengths and has been applied to a wide range of ground-based and spacecraft observations. More recently, \citeA{melin20} extended these capabilities into the ultraviolet in order to model Cassini UVIS observations at 150--190 nm.

At wavelengths \textgreater 164 nm, Jupiter's spectrum is dominated by Rayleigh-scattered sunlight. NEMESIS calculates multiple scattering by using a matrix operator approach \cite{plass73}. This scattered sunlight is absorbed by molecules in Jupiter's atmosphere, such as C\textsubscript{2}H\textsubscript{2}. Light can also be absorbed and/or scattered by aerosols in Jupiter's atmosphere. For a given set of vertical temperature, abundance and aerosol profiles, viewing geometry and incident solar spectrum, NEMESIS calculates a synthetic top-of-atmosphere spectrum. This synthetic spectrum is compared to the observed spectrum and the atmospheric profiles are iteratively adjusted to fit the observed spectrum, following an optimal estimation approach~\cite{rodgers00}. The final fit minimizes the chi-square statistic between the observations and the model, taking into account both the noise on the observations and an estimated 10\% error in the radiative transfer calculations.

Raman scattering, in which photons are scattered inelastically, is not currently included in NEMESIS. At wavelengths greater than 210 nm, Raman scattering can have a significant impact on the reflectance spectrum, contributing 3--6\% to the continuum and producing multiple identifiable spectral features~\cite{betremieux99}. In contrast, at less than 210 nm, Raman scattering does not produce clear spectral features and contributes 2--3\% to the continuum. This is a similar size to the noise on the observations, and forms part of the assumed 10\% error in the radiative transfer calculations.

In this paper, we use the atmosphere ``Model C'' described in \citeA{moses05} as the a priori atmospheric model. In some retrievals, we also include NH\textsubscript{3} and PH\textsubscript{3}, which are not present in the stratosphere-focused Model C. The a priori NH\textsubscript{3} and PH\textsubscript{3} vertical profiles are obtained from Cassini CIRS observations \cite{fletcher16} and assume an abundance of zero above the tropopause. The sources of ultraviolet absorption cross-sections and the solar spectrum are described in \citeA{melin20} and both were smoothed to match the spectral resolution of the Juno UVS wide slit. For simplicity, we assume that the aerosols in the stratosphere consist of a purely absorbing grey (spectrally-flat) haze; this assumption provides a good fit to the data and we do not have enough spectral information to support a more complex haze model. Even with a grey haze, the haze location can affect the spectral shape and in the following section we discuss this effect on the retrieved abundances. 

A very limited number of parameters are allowed to vary in these NEMESIS retrievals: the abundances of a subset of atmospheric gases (each via a single scaling factor applied to the a priori vertical profile) and the optical thickness of the stratospheric haze. The red line in Figure~\ref{fig:spectrum} shows the fit that was obtained for the equatorial spectrum using only three parameters: the C\textsubscript{2}H\textsubscript{2} and C\textsubscript{2}H\textsubscript{6} abundances and the opacity of an extended stratospheric haze layer. In this example fit, NH\textsubscript{3} and PH\textsubscript{3} were not included in the model. Even with such a limited parameter space, we are able to achieve a good fit to the data. In the following section, we perform retrievals on the 164--200 nm part of the spectrum, and therefore C\textsubscript{2}H\textsubscript{6} absorption has no impact and is not allowed to vary. We initially allow only C\textsubscript{2}H\textsubscript{2} and the haze opacity to vary, and then subsequently allow for NH\textsubscript{3} and PH\textsubscript{3} variability in order to confirm that they do not affect the qualitative results. 

\subsection{Latitudinal retrievals}
\label{sec:retrievals}

As shown in \citeA{melin20}, at wavelengths 164--200 nm, ultraviolet photons are primarily reflected from the 5--50 mbar region, with longer wavelengths probing slightly deeper in the atmosphere. Even though we assume a spectrally flat haze, if the location of the haze intersects with this region of peak sensitivity, it is possible for the haze to affect the spectral shape. Instead, if we assume that the haze is located much higher in the atmosphere, well above the region of peak sensitivity, then increasing its opacity simply scales down the spectrum without impacting its shape. This in turn means that the spectral shape variability shown in Figure~\ref{fig:contour} is almost entirely due to C\textsubscript{2}H\textsubscript{2} variability.

\begin{figure}
\centering
\includegraphics[width=10cm]{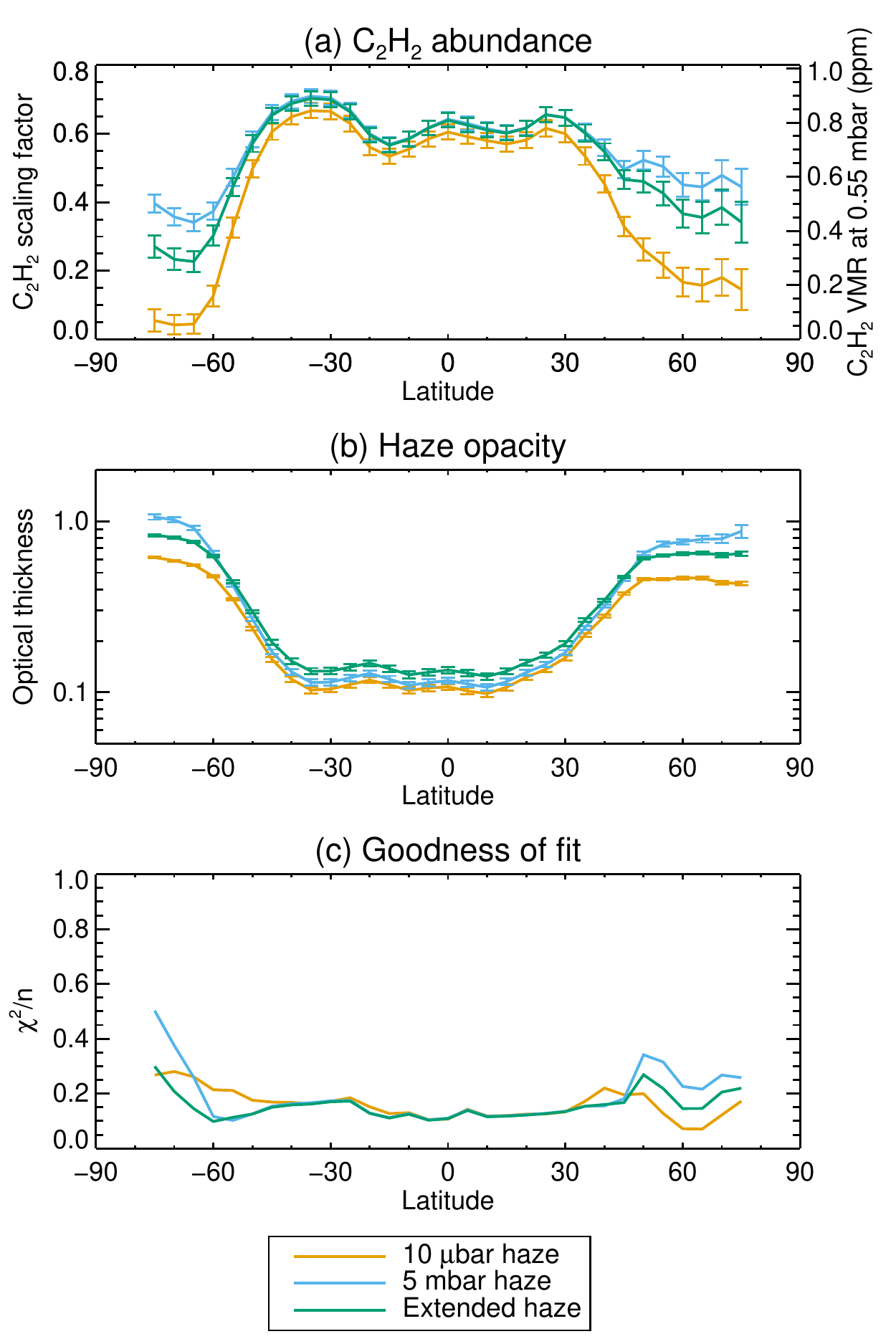}
\caption{The results of the latitudinal retrievals for three different cloud models. (a) shows the retrieved C\textsubscript{2}H\textsubscript{2} abundance as a function of latitude. The left axis shows the abundance in terms of the retrieved scaling factor relative to Model C in \citeA{moses05} and the right axis shows how this scaling factor converts into a relative abundance at 5 mbar. (b) shows the retrieved haze opacity and (c) shows the reduced chi-square statistic.}
\label{fig:c2h2}
\end{figure}

To explore this simple case, we performed retrievals at all latitudes in which only two parameters were allowed to vary: the scaling factor of the C\textsubscript{2}H\textsubscript{2} vertical profile and the opacity of a compact haze layer located at 10 \textmu bar, well above the region of peak sensitivity. The results of this retrieval are shown in orange in Figure~\ref{fig:c2h2}. Figure~\ref{fig:c2h2}(a) shows the C\textsubscript{2}H\textsubscript{2} scaling factor, which is defined relative to the \citeA{moses05} Model C vertical profile i.e., a scaling factor of 1.0 is the same as the \citeA{moses05} vertical profile while a scaling factor of 0.5 is half as abundant at all pressure levels. In Model C, the C\textsubscript{2}H\textsubscript{2} density peaks at 0.55 mbar and it is therefore this part of the atmosphere that has the strongest influence on the C\textsubscript{2}H\textsubscript{2} retrievals, although the spectrum is sensitive to the total column abundance down to the 5--50 mbar region. As expected from Figure~\ref{fig:contour}, the orange line in Figure~\ref{fig:c2h2}(a) shows that the  C\textsubscript{2}H\textsubscript{2} abundance decreases rapidly at high latitudes; by $\pm$60$^{\circ}$, the abundance is approximately a quarter of the low-latitude value. Figure~\ref{fig:c2h2}(b) shows the retrieved haze opacity, which increases towards the poles as expected and Figure~\ref{fig:c2h2}(c) shows the reduced chi-square statistic, which shows that this simple two-parameter retrieval can achieve a good fit ($\chi^2/n$ \textless 1) at all latitudes.

If the haze is moved deeper in the atmosphere, into a region where it intersects with the region of peak sensitivity, then changing the haze opacity can also alter the spectral shape. As shown in \citeA{melin20}, the longer wavelength part of the spectrum probes slightly deeper in the atmosphere, both due to an absence of stratospheric absorbers and because of decreased Rayleigh scattering. A haze layer placed within the region of peak sensitivity therefore absorbs a higher fraction of the light at long wavelengths, since at shorter wavelengths, some of the sunlight is reflected before it reaches the haze layer. This can introduce a `tilt' in the spectrum, which can be compensated for by increasing the C\textsubscript{2}H\textsubscript{2} abundance i.e., it introduces a degeneracy between the haze opacity and the C\textsubscript{2}H\textsubscript{2} abundance. This can be seen from the blue lines in Figure~\ref{fig:c2h2}, which show the retrieval results when a compact haze layer at 5 mbar is used; towards the poles, where the opacity is higher, the blue line has a higher C\textsubscript{2}H\textsubscript{2} abundance than the orange line. This is particularly true in the northern hemisphere, where the abundance ticks upward to reach the low-latitude levels. However, it should be noted that at these most northern latitudes, the 5-mbar haze model produces a worse fit to the data than the 10-\textmu bar model, and continuing to move the compact haze deeper into the atmosphere continues to worsen the fit.

These two compact haze models represent the two extremes of the possible haze location; high enough that it does not overlap with the region of peak sensitivity and the deepest possible level that still allows for a good quality fit to the data. In reality, neither of these models is physically realistic, as the stratospheric haze is likely to be extended over a large pressure range. As there is haze opacity within the region of peak sensitivity and also at higher altitudes, the true C\textsubscript{2}H\textsubscript{2} distribution likely lies between the blue and orange lines in Figure~\ref{fig:c2h2}(a). To explore this, we created an extended haze model with constant density in the 100--1 mbar range, approximated from the aerosol model described in \citeA{wong03}. The latitudinal retrieval results using this model are shown in green in Figure~\ref{fig:c2h2}. As expected, the retrieved C\textsubscript{2}H\textsubscript{2} abundances lie between the results from the other two models. In addition, the quality of the fit is better at almost all latitudes. We note that despite the very simple haze parametrization, the retrieved latitudinal haze opacities shown in Figure~\ref{fig:c2h2}(b) are very similar to the asymmetric latitudinal haze distribution obtained by \cite{zhang13} from Cassini ISS and ground-based near-infrared observations. 

Thus far, our retrievals have been limited to C\textsubscript{2}H\textsubscript{2} and the haze opacity, and tropospheric NH\textsubscript{3} and PH\textsubscript{3} have not been included in the atmospheric model. In order to understand the impact that they may have on the retrieved C\textsubscript{2}H\textsubscript{2} distribution, we conducted another series of retrievals using the extended haze model where NH\textsubscript{3} and PH\textsubscript{3} were included and were also allowed to vary via a scaling parameter. This inclusion had minimal impact on the retrieved C\textsubscript{2}H\textsubscript{2} abundances (although it does increase the error bars) and did not improve the quality of the fit to the data.

%% ------------------------------------------------------------------------ %%
%
%  DISCUSSION + CONCLUSIONS
%
%% ------------------------------------------------------------------------ %%

\section{Discussion and conclusions}
\label{sec:discussion}

In this paper, we use ultraviolet reflected sunlight observations of Jupiter to determine how the planet's stratospheric C\textsubscript{2}H\textsubscript{2} abundance varies with latitude. We performed latitudinal retrievals where we allowed the C\textsubscript{2}H\textsubscript{2} abundance to vary via a scaling factor and found that this scaling factor peaks at low latitudes and decreases towards the poles. In these retrievals, we assume that all latitudes have the same relative vertical distribution of C\textsubscript{2}H\textsubscript{2}; in reality this may vary and the UVS data are not capable of vertically resolving the abundance profile. However, we carefully selected our Juno UVS data such that all of the latitudinal spectra have the same solar and observational geometry and therefore probe down to the same region of the atmosphere. The scaling factor is therefore representative of the total C\textsubscript{2}H\textsubscript{2} abundance above the 5--50 mbar level. The precise latitudinal distribution depends on the assumptions made about the stratospheric haze altitude, but the high latitude abundances are approximately 2--4 times lower than the equatorial abundances.

There have been several papers that have studied Jupiter's meridional distribution of C\textsubscript{2}H\textsubscript{2} using infrared observations, from both spacecraft \cite{nixon07,nixon10,zhang13b} and groundbased telescopes \cite{fletcher16,melin18}. High resolution infrared observations have the advantage in being able to separate individual emission lines and can be used to retrieve the vertical profile of a molecular species. However, ultraviolet observations have the advantage that they are not dependent on the temperature profile in the stratosphere, a significant difficulty in infrared analyses \cite<e.g.,>{fletcher16}. The results presented in this paper are the first full-planet meridional retrievals of C\textsubscript{2}H\textsubscript{2} from ultraviolet observations and they build on the work of \citeA{melin20} which presents low-latitude retrievals from Cassini UVIS. 

\begin{table}

\begin{center}
\begin{tabular}{|
>{\raggedright\arraybackslash}p{0.9cm} | 
>{\raggedright\arraybackslash}p{1.6cm} |
>{\raggedright\arraybackslash}p{1.6cm} |
>{\raggedright\arraybackslash}p{5.0cm} |
>{\raggedright\arraybackslash}p{2.2cm} |
} 
\hline
 \bf{Year} & \bf{Jupiter season} & \bf{Data source} & \bf{Meridional trend} & \bf{Reference} \\ \hline
1979 & Northern fall & Voyager IRIS & Constant with latitude at 0.1--7 mbar & \citeA{nixon10} \\ \hline
2000 & Northern summer & Cassini CIRS & Decreases towards poles by a factor of $\sim$4 at 5 mbar, clear north-south asymmetry (maximum at 20${^\circ}$N) & \citeA{nixon07} \\ \hline
%2000 & Northern summer & Cassini UVIS & Approximately constant between 40${^\circ}$N and 40${^\circ}$S (no high-latitude coverage) & \citeA{melin20} \\ \hline
2014 & Northern fall & IRTF TEXES & Decreases towards poles by a factor of $\sim$2 at 5 mbar, some north-south asymmetry but less than in Cassini CIRS data & \citeA{fletcher16} \\ \hline
2013--2017 & Northern fall & IRTF TEXES & Decreases towards poles by factors of 2--3 at 1 mbar, north-south asymmetry decreases over 4 years of observations, symmetric by 2017 & \citeA{melin18} \\ \hline
2016--2020 & Northern winter & Juno UVS & Decreases towards poles by a factor of 2--4 at pressures less than 5--50 mbar, no clear north-south asymmetry & This paper \\ \hline
\end{tabular}
\end{center}
\caption{Comparison of full-planet meridional C\textsubscript{2}H\textsubscript{2} studies.}
\label{tab:studies}
\end{table}

Table~\ref{tab:studies} describes the results from previous studies of Jupiter's meridional C\textsubscript{2}H\textsubscript{2} distribution. The decrease of C\textsubscript{2}H\textsubscript{2} by a factor of up to 4 at high latitudes is a trend that was seen in three of the four previous full-planet studies \cite{nixon07,fletcher16,melin18}. This clear agreement between observations from different spectral regions is reassuring. This result also agrees with the expectation from solar insolation rates; near the equator, the UV flux is higher, allowing more C\textsubscript{2}H\textsubscript{2} to be generated from the UV photolysis of CH\textsubscript{4}. Including the effect of horizontal mixing complicates this picture and can act to flatten the distribution. The horizontal mixing efficiency has previously been estimated by tracking the post-impact spread of SL9-related species over decades, and by studying the hydrocarbon distribution using coupled chemistry-transport models; the constraints provided by these observations are limited in pressure and latitude, but assuming a reasonable range of horizontal mixing processes has been shown to still result in a decrease towards the poles in the C\textsubscript{2}H\textsubscript{2} abundance \cite{hue18b}.

The exception to this latitudinal trend is the reanalysis of Voyager IRIS data; while an earlier preliminary analysis suggested that C\textsubscript{2}H\textsubscript{2} decreases at high latitude in the Voyager IRIS observations \cite{maguire84}, \citeA{nixon10} reanalyzed the data using the same methods used for the Cassini CIRS data and found that the abundance was approximately constant with latitude (up to $\pm$60$^{\circ}$). \citeA{nixon10} suggested that this could be due to seasonal effects from Jupiter's eccentric orbit, although subsequent observations during the same season (northern fall) three Jovian years later \cite{fletcher16,melin18} do show a decrease towards the poles and modeling work from \citeA{hue18b} shows that there should be minimal seasonal effects on the hydrocarbon abundances at these depths.

In addition to the high-latitude decrease, the studies listed in Table~\ref{tab:studies} also discuss the (a)symmetry of the meridional distribution at lower latitude. In 2000, \citeA{nixon07} found that the peak C\textsubscript{2}H\textsubscript{2} abundance was at 20$^{\circ}$N rather than at the equator. \citeA{fletcher16} observed a similar asymmetry in 2014 and \citeA{melin18} observed that this asymmetry decreased over the 2013--2017 time period. \citeA{melin18} suggested that the cause for this asymmetry could be related to periods of strong wave activity in Jupiter's Northern Equatorial Belt or other dynamic tropospheric events that might lead to material or wave transfer of energy with altitude. Our observations in this paper were obtained during 2016--2020 and do not show any strong north-south asymmetry within the low-latitude region. The individual latitudinal distributions shown in Figure~\ref{fig:c2h2} do show some asymmetries at high latitudes, but when we consider all four model runs together, this asymmetry falls within the error bars.

In this paper, we focused exclusively on meridional trends, and we excluded auroral regions from the highest-latitude spectra, due to the complexity of analyzing spectra that combine auroral emission and reflected sunlight. The meridional infrared studies we discuss similarly avoided regions of the planet with auroral emission \cite{nixon07,melin18}. However, within the polar region of the planet, the hydrocarbon distribution appears to vary strongly with latitude; using infrared observations, \citeA{sinclair17b,sinclair18} found that there is an enrichment in C\textsubscript{2}H\textsubscript{2} within the auroral ovals, relative to the surrounding quiescent longitudes. As a polar-orbiting spacecraft, Juno provides excellent views of Jupiter's poles. In the future, we will therefore seek to expand this study and use Juno UVS observations to map the polar C\textsubscript{2}H\textsubscript{2} distribution. 

\section*{Acknowledgements}

We are grateful to NASA and contributing institutions, which have made the Juno mission possible. This work was funded by NASA's New Frontiers Program for Juno via contract with the Southwest Research Institute. HM and LNF were supported by a European Research Council Consolidator Grant (under the European Union's Horizon 2020 research and innovation programme, grant agreement No 723890) at the University of Leicester. PGJI acknowledges the support of the United Kingdom’s Science and Technology Facilities Council. BB is a Research Associate of the Fonds de la Recherche Scientifique - FNRS. 

\section*{Data Availability Statement}

The Juno UVS data used in this paper are archived in NASA's Planetary Data System Atmospheres Node: https://pds-atmospheres.nmsu.edu/PDS/data/jnouvs\_3001 \cite{trantham14}. The data used to produce the figures in this paper are available in \citeA{giles21_data}. The NEMESIS radiative transfer and retrieval tool is available from \citeA{irwin20_nemesis}.

%\bibliography{main.bib}

\end{document}